\documentstyle[11pt,IAUS212,twoside,epsf]{article}

\def\edcomment#1{\iffalse\marginpar{\raggedright\sl#1\/}\else\relax\fi}
\reversemarginpar

\def\plotonea#1{\centering \leavevmode
\epsfxsize=.6\textwidth \epsfbox{#1}}
\def\plotoneb#1{\centering \leavevmode
\epsfxsize=.75\textwidth \epsfbox{#1}}
\def\plottwoa#1#2{\centering \leavevmode
\epsfxsize=.35\textwidth \epsfbox{#1} \hfil
\epsfxsize=.35\textwidth \epsfbox{#2}}
\def\plottwob#1#2{\centering \leavevmode
\epsfxsize=.35\textwidth \epsfbox{#1} \hfil
\epsfxsize=.6\textwidth \epsfbox{#2}}

\begin{document}
\vspace*{1cm}
\title{Ring Nebulae around Massive Stars throughout the HR Diagram}
 \author{You-Hua Chu}
\affil{Astronomy Department, University of Illinois, 1002 W. Green 
Street, Urbana, IL 61801, USA}

\begin{abstract}
Massive stars evolve across the HR diagram, losing mass along the 
way and forming a variety of ring nebulae.  During the main 
sequence stage, the fast stellar wind sweeps up the ambient 
interstellar medium to form an interstellar bubble.  After a massive 
star evolves into a red giant or a luminous blue variable, it loses 
mass copiously to form a circumstellar nebula.  As it evolves 
further into a WR star, the fast WR wind sweeps up the previous 
mass loss and forms a circumstellar bubble.  Observations of ring 
nebulae around massive stars not only are fascinating, but also are
useful in providing templates to diagnose the progenitors of 
supernovae from their circumstellar nebulae.  In this review, I will
summarize the characteristics of ring nebulae around massive stars
throughout the HR diagram, show recent advances in X-ray observations 
of bubble interiors, and compare supernovae's circumstellar nebulae 
with known types of ring nebulae around massive stars.
\end{abstract}

\section{Introduction}

Since Johnson \& Hogg (1965) reported the first three ring
nebulae around Wolf-Rayet (WR) stars, nebulae of various
shapes, sizes, and ionization conditions have been observed 
around massive stars of different spectral types, such as 
luminous blue variables (LBVs), blue supergiants (BSGs), and 
red supergiants (RSGs).
As these various spectral types in the upper part of the
Hertzsprung-Russell (HR) diagram are strung together by the 
evolutionary tracks of massive stars, their surrounding
nebulae must be evolutionarily related.
This relationship was illustrated clearly by Garc\'{\i}a-Segura,
Langer, \& Mac Low (1996) and Garc\'{\i}a-Segura, Mac Low, \& 
Langer (1996) in their hydrodynamic simulations of the formation
of WR ring nebulae taking into account the evolution and mass
loss history of the central stars.

In this review, I will cover three topics.  First, I will 
present a gallery of ring nebulae around massive stars 
throughout the HR diagram and compare them to Garc\'{\i}a-Segura 
et al.'s framework of hydrodynamic evolution of circumstellar 
gas around massive stars.  Second, I will report an X-ray view 
of the hot gas in the interiors of bubbles blown by massive
stars, using ROSAT, ASCA, Chandra, and XMM-Newton observations.
Finally, I will compare the circumstellar nebulae observed around 
young supernovae (SNe) to those 
around massive stars and use the nebular properties to diagnose
the spectral types of SN progenitors.

\section{Gallery of Ring Nebulae around Massive Stars}

The hydrodynamic models of ring nebulae around massive stars 
by Garc\'{\i}a-Segura et al.\ considered two possible stellar 
evolution tracks from the main sequence (MS) to the WR phase:
MS O $\rightarrow$ LBV $\rightarrow$ WR, for a 60 M$_\odot$ star;
MS~O $\rightarrow$ RSG $\rightarrow$ WR, for a 35 M$_\odot$ star.
Stars at these different evolutionary stages expel stellar 
material in the following ways: MS O stars possess tenuous, fast 
winds with terminal velocities of 1,000--2,000 km~s$^{-1}$; 
RSGs have copious, slow winds with wind velocities typically a 
few 10's km~s$^{-1}$; LBVs have outbursts or winds at modest 
velocities that depend on the stellar luminosities; WR stars 
have the most powerful stellar winds with both large mass loss 
rates, $\sim10^{-5}$ M$_\odot$~yr$^{-1}$, and high wind 
velocity, up to 3,000 km~s$^{-1}$.  These stellar winds
interact with the ambient medium and form ring nebulae with
distinct characteristics for each type of central stars.

\subsection{Main Sequence O Stars -- Interstellar Bubbles}

The fast stellar wind of a MS O star sweeps up the ambient 
interstellar medium (ISM) to form an {\it interstellar bubble}
(Weaver et al.\ 1977), which consists of a dense shell of 
interstellar material.  Intuitively, we would expect around
most O stars an interstellar bubble similar to the Bubble 
Nebula (NGC 7635) to be visible; however, hardly any O stars 
in H\,{\sc ii} regions have ring nebulae, suggesting that
these interstellar bubbles are rare.  This puzzle is 
recently solved by observations of the
young H\,{\sc ii} region N11B in the Large Magellanic Cloud (LMC).
While no ring nebulae can be identified morphologically, long-slit
high-dispersion spectra have revealed expanding shells around
single or groups of O stars (see Fig.~1a,b; Naz\'e et al.\ 2001).  
Their expansion velocities, typically 10--15 km~s$^{-1}$, are not 
much higher than the isothermal sound velocity of the 10$^4$ K 
ionized gas in the H\,{\sc ii} region, and consequently cannot 
produce strong shock compression.  Lacking large density 
enhancements, these slowly expanding interstellar bubbles do not 
have visible ring nebulae.

\begin{figure}
\plotone{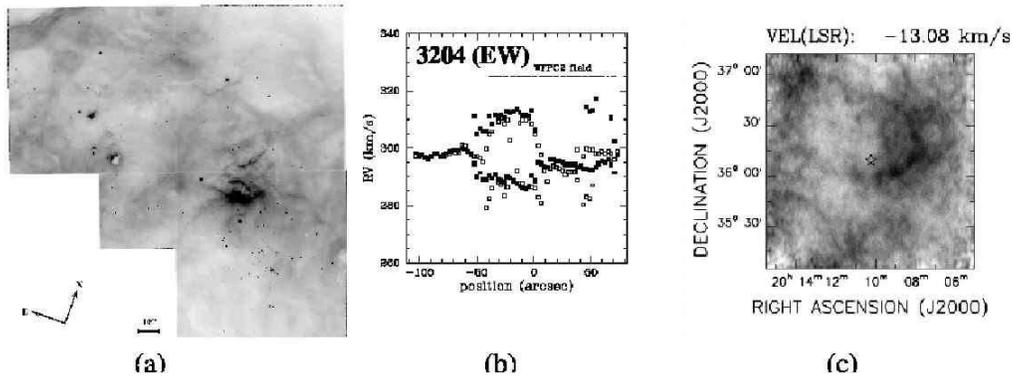}
\caption{(a) HST WFPC2 H$\alpha$ image of interstellar bubbles 
in N11; (b) velocity-position plot of an interstellar bubble in
N11B - from Naz\'e et al.\ 2001; (c) an H\,{\sc i} interstellar 
bubble blown by the MS O progenitor of WR 134 - from Gervais \& 
St-Louis 1999.}
\end{figure}

Interstellar bubbles blown by MS O stars may become 
morphologically identifiable at later evolutionary stages
when the central stars can no longer ionize the bubble
shells and the surrounding ISM.  A 10 km~s$^{-1}$ expanding 
shell in a 100~K neutral H\,{\sc i} medium (isothermal sound 
velocity $\sim$ 1 km~s$^{-1}$) is highly supersonic, thus 
can generate strong compression and produce an identifiable 
shell morphology.  Such recombined interstellar bubbles
have been observed as H\,{\sc i} shells around WR stars 
(Cappa 2002, and references therein).  
A beautiful H\,{\sc i} shell around WR 134 in our Galaxy
reported by Gervais \& St-Louis (1999) is reproduced in
Fig.~1c.

\subsection{Red Supergiants -- Circumstellar Nebulae}

Evolved massive stars at the RSG phase lose mass via slow 
winds and form circumstellar nebulae.  Because of the high 
dust content, these circumstellar nebulae are best observed
through the optical continuum scattered by dust or the thermal
IR continuum emitted by dust.  This is illustrated by VY CMa 
(M5e~Ia) in Fig.~2.  The optical reflection nebula of VY CMa,
visible to a radius of 9000 AU (for a distance of 1.5 kpc),
is asymmetric about the central star.  The presence of bright 
arcs indicates that localized ejection of stellar material 
may have occurred on the stellar surface (Smith et al.\ 2001).
More details on mass loss and circumstellar nebulae of
RSGs are reviewed by Humphreys (2002).

\begin{figure}
\plotonea{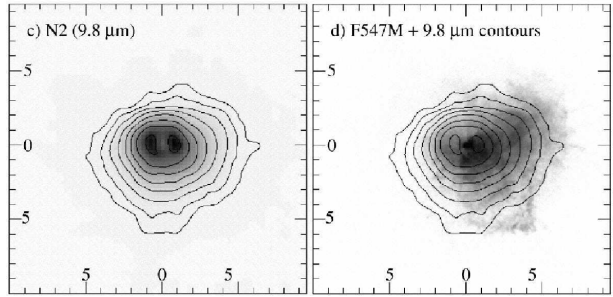}
\caption{The circumstellar nebula of VY CMa;
left: 9.8 $\mu$m image of VY CMa (M5e Ia).  
Right: HST WFPC2 F547M image of VY CMa overlaid by 
9.8 $\mu$m contours.  This figure is adopted from 
Smith et al.\ (2001).}
\end{figure}

\subsection{Luminous Blue Variables -- Circumstellar Bubbles}

The name LBV has been used for both the extreme case of $\eta$ Car
and the less luminous S Doradus variables.  The Ofpe/WN9 stars
have also been loosely called LBVs.  The circumstellar nebulae of
LBVs have been reviewed by Nota et al.\ (1995).  Using the
heavy element abundances of LBV nebulae, Smith et al.\ (1998) 
suggest that LBVs have gone through a brief RSG phase, while 
Lamers et al.\ (2001) demonstrate that the nebulae are ejected 
during the blue supergiant (BSG) phase and that the LBVs have 
never gone through a RSG phase.  

Most circumstellar nebulae around LBVs are small, $<$2 pc in 
diameter, expanding shells with V$_{\rm exp}$ of few 10's
km~s$^{-1}$ (Nota et al.\ 1995; Weis 2002).
Many LBV nebulae in the Galaxy and in the Magellanic Clouds
have been studied recently by, e.g., Nota et al.\ (1996, 1997),
Pasquali et al.\ (1997, 1999), and Weis (2002, and
references therein).  A large compilation of images of LBV 
nebulae were presented in the poster by Weis in this meeting.
Fig.~3 shows the intriguing runaway LBV S119 and its nebula 
in the LMC; the interstellar absorption lines in the FUSE 
spectrum of S119 have established unambiguously that it is 
in the LMC (Danforth \& Chu 2001).

\begin{figure}
\plotone{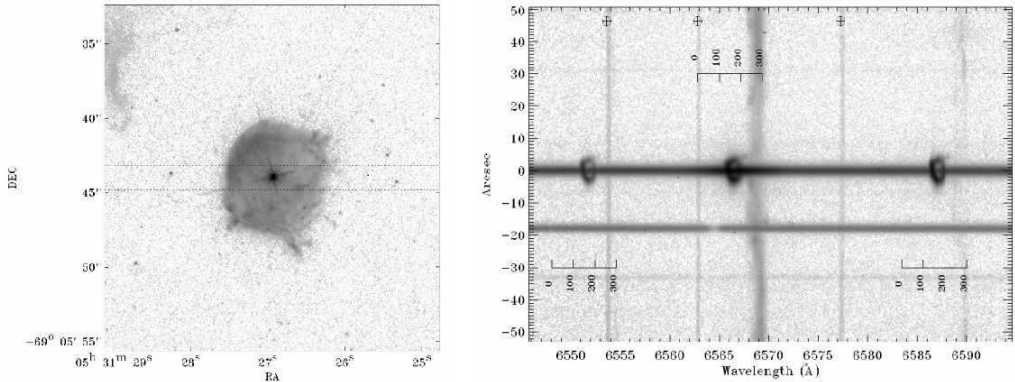}
\caption{The runaway LBV S119 and its circumstellar nebula.
Left: HST WFPC2 image in the H$\alpha$ line.  Right: long-slit 
echellogram of the H$\alpha$ and [N II] lines.  The ionized ISM
of the LMC is detected at heliocentric velocities of 220-290
km~s$^{-1}$ along the entire slit length.  The circumstellar 
nebula of S119 showns an expanding shell structure centered 
at $\sim$160 km~s$^{-1}$ with a line splitting of 46 km~s$^{-1}$.
This figure is adopted from Danforth \& Chu (2001).}
\end{figure}

\subsection{Wolf-Rayet Stars -- Interstellar/Circumstellar Bubbles}

In the last decade, sensitive surveys of WR ring nebulae have 
been made using CCD cameras with interference filters for the 
Galaxy (Miller \& Chu 1993; Marston et al.\ 1994a,b; Marston 
1997) and the Magellanic Clouds (Dopita et al.\ 1994).
The identification criteria in these surveys are similar to
those used in the 70's and 80's.
Based on nebular dynamics, Chu (1981) find three types of WR ring
nebulae: R-type nebulae are not dynamically shaped by the WR stars,
W-type nebulae are bubbles blown by the WR stars, and E-type nebulae
are ejecta nebulae. 

Garc\'{\i}a-Segura et al.\ (1996a,b) show that a WR star is
surrounded by an inner circumstellar bubble and an outer 
interstellar bubble.  The interstellar bubble was blown by the 
progenitor of the WR star at the MS phase, while the circumstellar
bubble is blown by the WR star in the circumstellar nebula 
ejected by the progenitor of the WR star at the RSG or LBV phase.
Both the E-type and W-type WR ring nebulae refer to the 
circumstellar bubbles.   If the circumstellar bubble is too
tenuous to be detected, then the outer interstellar bubble may
be ionized and become a R-type WR ring nebula.

It is thus clear that the three types of WR ring nebulae defined
by Chu (1981) are not mutually exclusive.  In particular, there
is no physical distinction between W-type and E-type nebulae.
Bearing in mind that the optically identified ring nebulae
around WR stars may refer to either the circumstellar bubbles 
or the interstellar bubbles, it is thus meaningless to make 
statistical statements about ``the percentage of WR stars in 
visible ring nebulae," unless distinction is made between
the circumstellar and interstellar cases.

WR ring nebulae have been studied in multiple wavelengths.
Using the differences in H$\alpha$ and [O\,{\sc iii}] images,
Gruendl et al.\ (2000) presented a morphological diagnostic 
for dynamical evolution of WR bubbles.  The most dramatic contrast
between H$\alpha$ and [O\,{\sc iii}] morphology is illustrated in
the HST WFPC2 images of NGC\,6888 in Fig.~4 (Moore, Hester, \&
Scowen 2000).  H\,{\sc i} 21-cm line observations have revealed
neutral gas shells of diameters a few 10's pc around WR stars;
the properties of these shells are consistent with those expected
in the interstellar bubbles blown by the MS progenitors (see
review by Coppa 2002).  

Observations of molecular gas associated with WR ring nebulae have 
yielded many intriguing results: (1) circumstellar CO associated
with WR\,16 (Marston et al.\ 1999) and possibly NGC\,6888 (Rizzo
et al.\ 2002); (2) complex molecules and shock-excited interstellar 
H$_2$ and NH$_3$ in NGC\,2359 (St-Louis et al.\ 1998; Rizzo, 
Mart\'{\i}n-Pintado, \& Henkel 2001); (3) interstellar molecular
gas around NGC\,3199 (Marston 2001, 2002).

\begin{figure}
\plotone{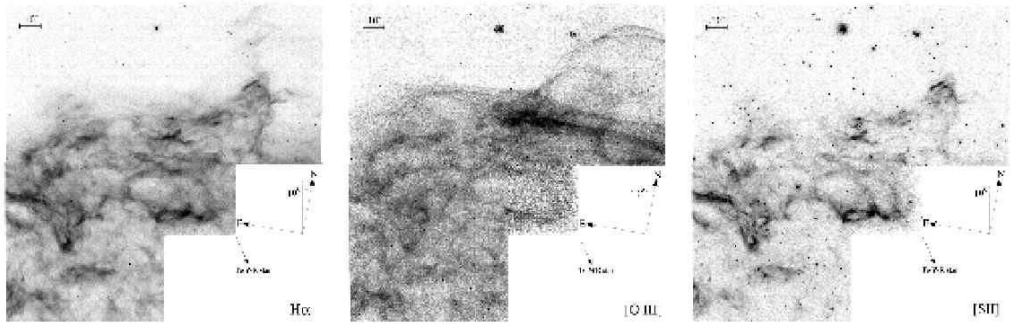}
\caption{HST WFPC2 images of NGC\,6888 in H$\alpha$, [O\,{\sc iii}]
and [S\,{\sc ii}] lines.  This figure is taken from Moore et al.\ 
(2000).}
\end{figure}

\subsection{Blue Supergiants and Massive Binaries -- Circumstellar
Nebulae}

The BSG phase has no unique meaning.  Any massive star that is blue
and luminous is called a BSG.  If it varies, then it is called a LBV.
The existence of a close binary companion can make the nature of a
BSG even more mysterious.  
A small number of circumstellar nebulae around BSGs have been 
identified.  Ironically, almost all of them have something peculiar 
about either the spectral properties of the stars or the physical 
structure of the nebulae.   Below are a few examples.

NGC\,6164-5 around 
HD\,148937 consists of a small (2.6 $\times$ 1.9 pc) S-shaped 
circumstellar nebula and a filamentary interstellar ring nebula 
($\sim$ 10 pc in diameter) inside a large dusty cavity structure
(Leitherer \& Chavarr\'{\i}a-K.\ 1987).  Clearly, HD\,148937 is
an evolved star, but its spectral type, O6.5f?p, is peculiar and
it shows well-marked emission in C\,{\sc iii} and N\,{\sc iii} lines
that are not seen in normal Of stars (Walborn 1972).  What is the
evolutionary status of HD\,148937?

Sk$-$69~202 (B3\,I), the progenitor of SN\,1987A, and Sher~25 
(B1.5\,I) are both BSGs and have circumstellar ring nebulae of 
similar size and morphology (Brandner et al.\ 1997a, b). 
The hourglass morphology of these two rings are uncommon among 
known ring 
nebulae around massive stars; in fact, no other ring nebulae have 
such a morphology.  Do these ring nebulae provide circumstellar
or circumstantial evidence for a similar evolutionary status
between these two BSGs?

Podsiadlowski, Joss, \& Hsu (1992) have proposed that Sk$-$69~202
is a product of a massive interacting binary; their model explains
the BSG spectral type of the star and the formation of the
equatorial ring in the circumstellar nebula.
Interestingly, a compact circumstellar nebula has been observed 
around the massive eclipsing binary RY Scuti (Smith et al.\ 1999);
another circumstellar nebula has been detected around the B[e]
supergiant R4 and is suggested to be ejected during a binary merger
process (Pasquali et al.\ 2000).
These circumstellar nebulae clearly have different formation
mechanisms from those around single massive stars.
The circumstellar nebula of Sher~25, resembling that around
Sk$-$69~202, most likely has a similar formation mechanism,
which then requires Sher~25 to involve a massive interacting 
binary.  Unless observations of Sher~25 can exclude this possibility,
the ring nebula of Sher~25 provides circumstellar evidence that
it is similar to Sk$-$69~202.

\begin{figure}
\plottwoa{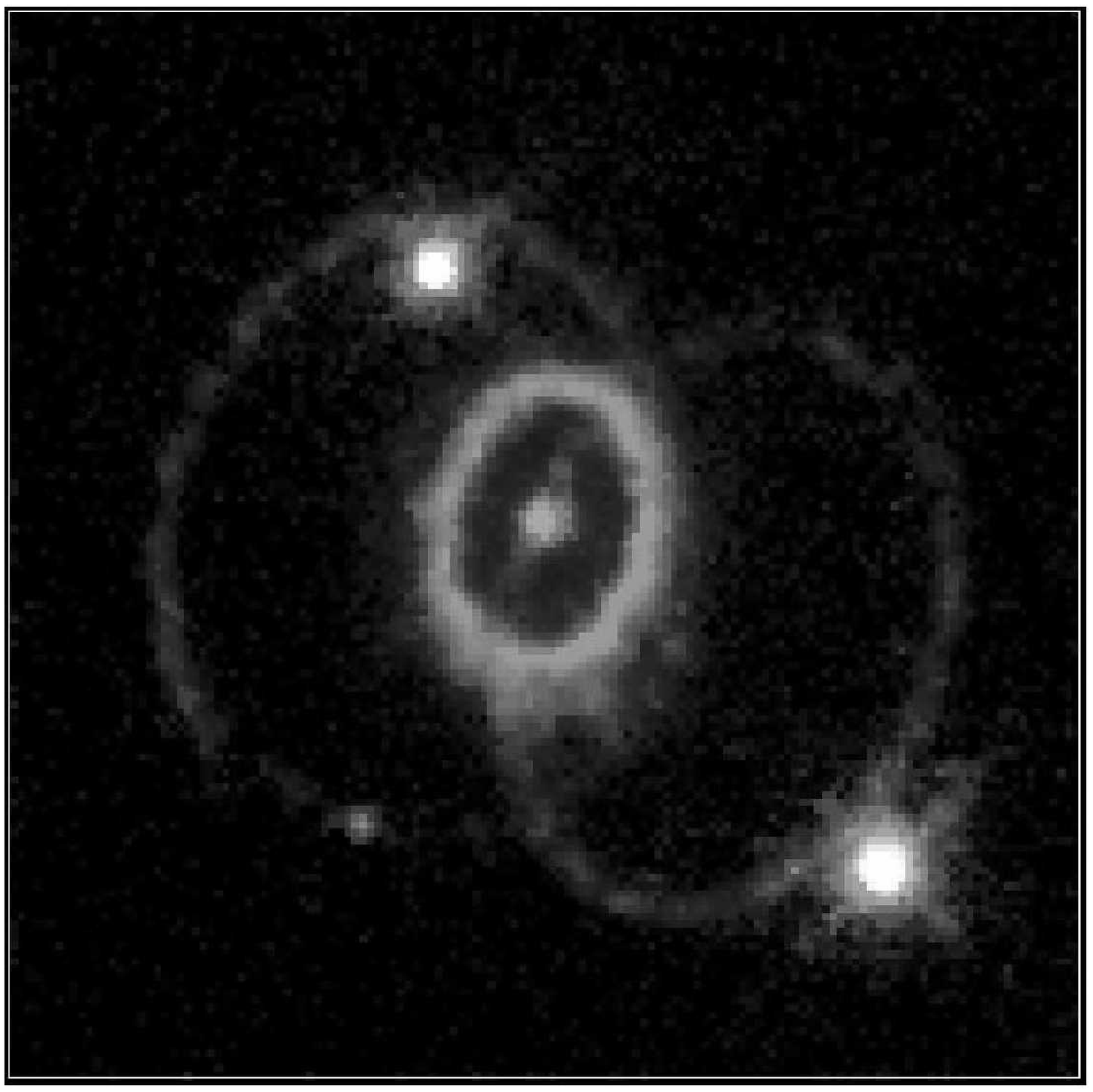}{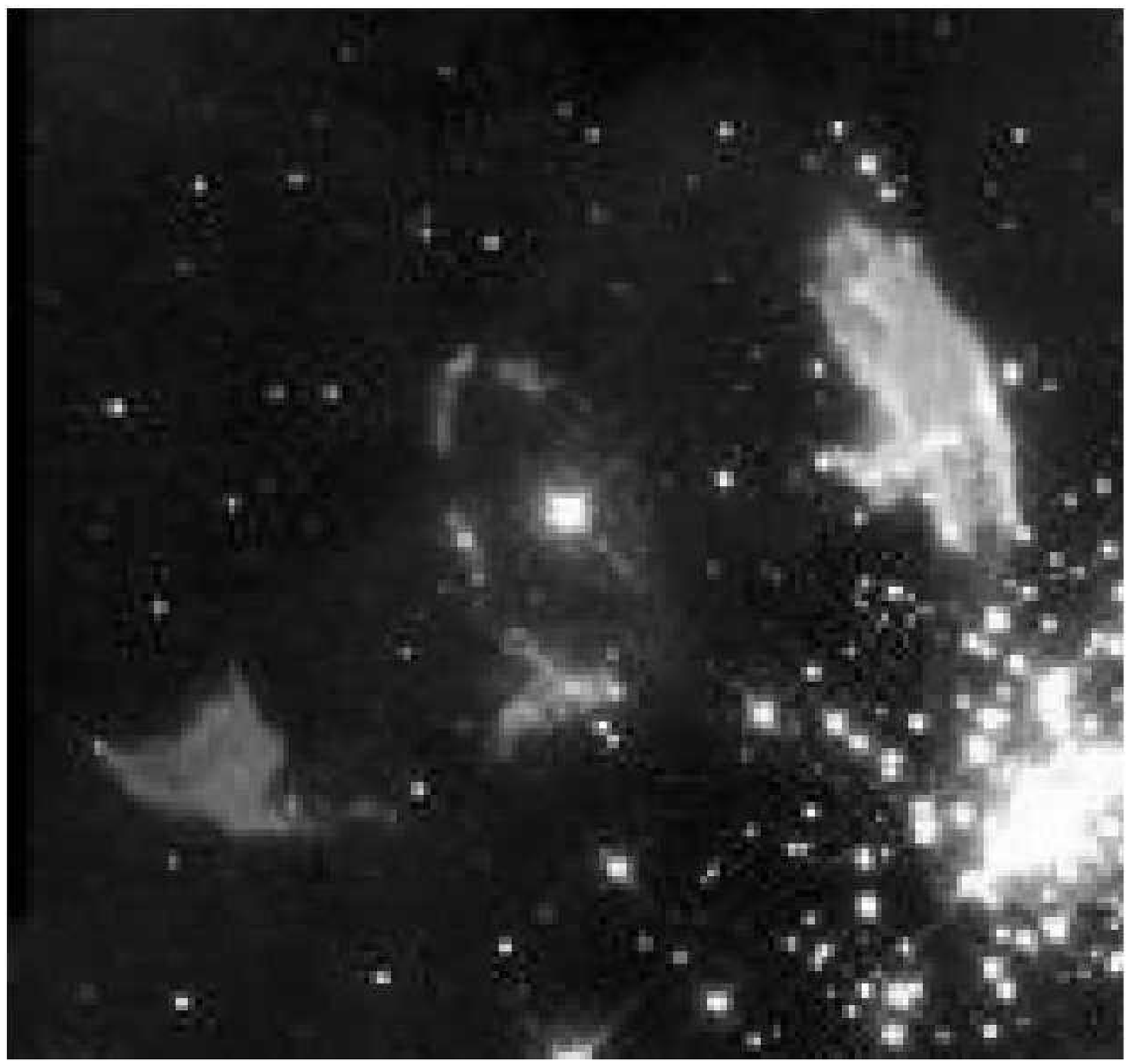}
\caption{Left: SN\,1987A and the circumstellar nebula of its 
B3\,I progenitor Sk$-$69~202.  Right: Sher~25, a B1.5\,I star,
and its circumstellar nebula.   Both inner rings are 0.4 pc in 
diameter.}
\end{figure}

\subsection{Observations vs. Expectations}

Observations of ring nebulae around single massive stars with known
evolutionary status generally agree with expectations: \\
(1) MS O stars blow interstellar bubbles of sizes $\sim$ a few 
    $\times$10 pc, \\
(2) RSGs are surrounded by circumstellar nebulae of sizes a few
     $\times$10$^3$ AU, \\
(3) LBVs have small circumstellar bubbles with radius $\le$1 pc, \\
(4) WR stars have larger circumstellar bubbles, a few pc in size, 
    surrounded by\\
\null~~~~ interstellar bubbles blown by their progenitors at MS stage. \\
The circumstellar nebulae, consisting of material ejected by the stars,
are N-enriched and can be diagnosed by a high [N~{\sc ii}]/H$\alpha$ 
ratio.  LBVs may evolve into WR stars, and the sizes of their
circumstellar bubbles (Chu, Weis, \& Garnett 1999) support this 
evolutionary sequence.

On the other hand, ring nebulae around massive stars with uncertain 
evolutionary status, such as BSGs, show complex structures that are 
not completely understood.  This is an area that needs a lot of future
work.  Ring nebulae around close massive binaries also need to be
studied.

\begin{figure}
\plottwob{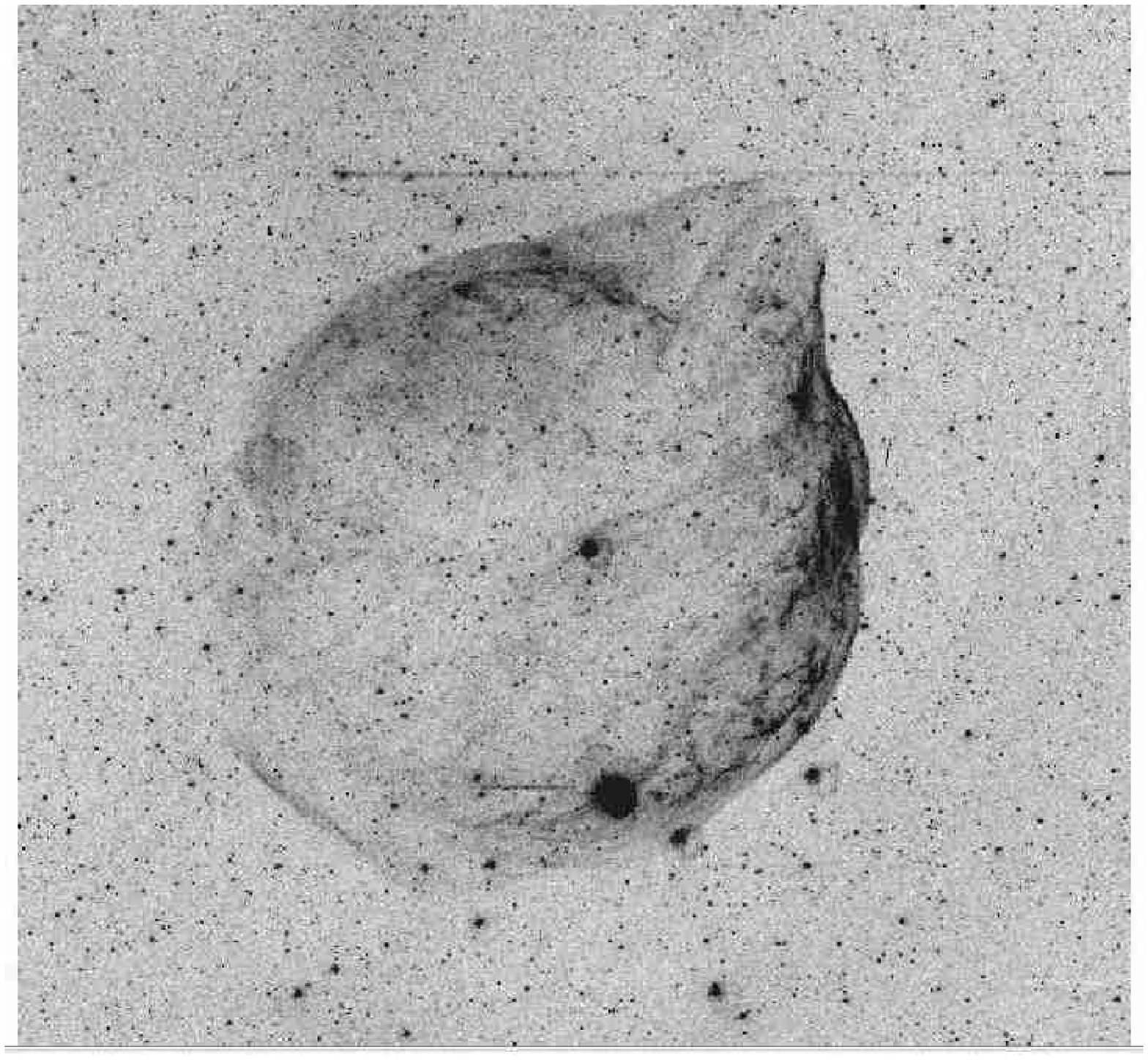}{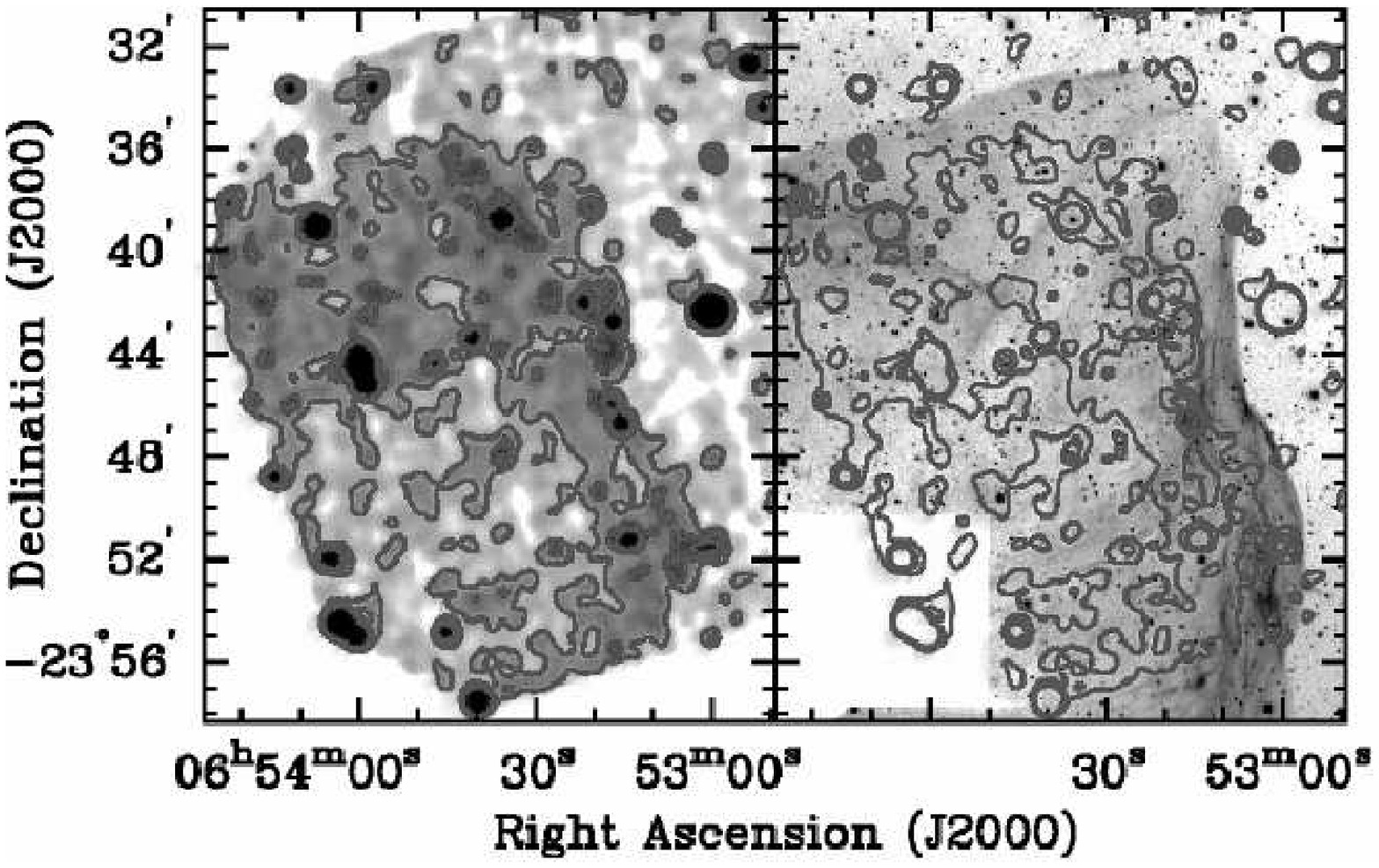}
\caption{[O\,{\sc iii}] (left), XMM X-ray (middle), and
X-ray contours over  [O\,{\sc iii}] images 
of S\,308.  The [O\,{\sc iii}] image was taken by Chris Smith 
with the Curtis Schmidt telescope.  The X-ray image covers only
the NW quadrant of S\,308.}
\end{figure}

\section{X-ray Views of Ring Nebulae}

Bubbles blown by massive stars ought to be filled by shocked
fast stellar wind at X-ray-emitting temperatures (Weaver et al.\
1977).  The first detection of diffuse X-ray emission from the
hot gas in a bubble interior was made by Einstein IPC observations
of NGC\,6888 (Bochkarev 1988).  ROSAT PSPC observations of several
WR ring nebulae detected only NGC\,6888 and S\,308 (Chu 1994).
The X-ray emission from NGC\,6888 is brighter toward the dense 
shell rim, and the X-ray spectrum indicates a plasma temperature 
of 1.6$\times$10$^6$ K (Wrigge, Wendker, \& Wisotzki 1994).
ROSAT PSPC observations of S\,308 is severely affected by the
occultation ring, making the extraction of X-ray spectrum difficult 
and unreliable (Wrigge 1999). ASCA observations of NGC\,6888 show an 
additional faint component at 8$\times$10$^6$ K (Wrigge et al.\ 1998).

The advent of Chandra and XMM-Newton finally makes it possible to 
study the distribution and physical conditions of the hot gas in 
bubble interiors.  To date, NGC\,6888 has been awarded a 100 ks
Chandra ACIS-S observation in Cycle 4, and S\,308 has recently 
been observed with XMM (Chu et al.\ 2002, in preparation).  
Figure 6 shows the
X-ray image of the northwest quadrant of S\,308.  For the
first time, the relative locations of the interior hot gas and
the cool shell are resolved!  A gap of 0.5 pc between the boundary
of hot gas and the leading edge of the cool, dense bubble shell
is observed; this gap would correspond to the interface region
where heat conduction takes place.

It is interesting to compare the hot interior of S\,308 to those
of planetary nebulae (PNe).  The progenitors of PNe have initial
masses up to $\sim$10 M$_\odot$, which corresponds to mid-B type
stars, which are qualified as ``massive stars."  PNe are believed
to be formed by the current fast stellar winds interacting with 
the former slow winds at the AGB phase (Kwok 1983), a mechanism
almost identical to that for the formation of a WR bubble with
a RSG progenitor (Garc\'{\i}a-Segura et al.\ 1996).  Diffuse
X-ray emission has been detected in four PNe (Chu, Guerrero,
\& Gruendl 2002, and references therein).  Figure 7 shows that
the hot gas in PN interiors has temperatures of 
2--3$\times$10$^6$ K, while S\,308  has the coolest interior,
only $\sim$1$\times$10$^6$ K.  These temperatures are much lower
than the post-shock temperature for a 1000--3000 km~s$^{-1}$ fast
wind.  Clearly, mixing with the cool nebular material has
taken place (e.g., Pittard et al.\ 2001a,b).  More X-ray
observations of ring nebulae around massive stars are needed to
constrain bubble models.

\begin{figure}
\plotone{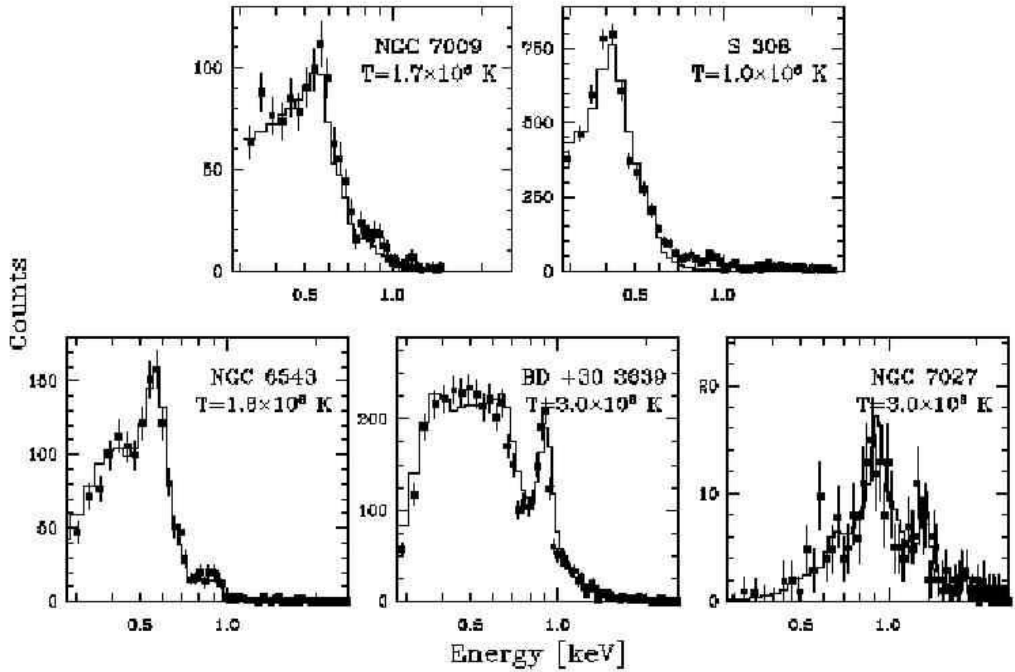}
\caption{X-ray spectra of shocked fast stellar winds.  The top
row are spectra taken with XMM-Newton, and the bottom row with
Chandra.  S\,308 is a WR bubble, while the others are planetary
nebulae.  The nebula name and plasma temperature derived from 
spectral fits are marked in each panel.}
\end{figure}

\section{Ring Nebulae after the Supernova Explosion}

One of the most intriguing questions about massive stellar evolution
is ``At what locations in the HR diagram do massive stars explode
as SNe?"  Only two SNe's progenitors have adequate data to estimate 
their spectral type: B3\,I for SN\,1987A (Walborn et al.\
1989) and K0\,I for SN\,1993J (Aldering, Humphreys, \& Richmond 1994).
Mass limits of SN progenitors have been estimated from photometric
measurements using pre-explosion images (Smartt et al. 2001, 2002;
Smartt 2002).  

An alternative way to estimate a SN progenitor's spectral type is 
through the physical properties of the circumstellar nebulae, which 
can be identified by narrow emission lines in high-dispersion
spectra of SNe.  The expansion of the nebula can be derived
from the widths of line profiles, and the nebular density and 
temperature can be derived from the
[O\,{\sc iii}] and [N\,{\sc ii}]
lines.   A variety of circumstellar nebulae have been observed.

SN\,1997ab and SN\,1997eg both showed a narrow 
P Cygni H$\alpha$ line, implying the existence of circumstellar
nebulae with densities $>$10$^7$ cm$^{-3}$ and expansion 
velocities 100--150 km~s$^{-1}$ (Salamanca et al.\ 1998; Salamanca,
Terlevich, \& Tenorio-Tagle 2002). These narrow P Cygni lines
disappeared in observations made in March 1999, indicating that
they are small (Gruendl et al.\ 2002). These circumstellar nebulae
cannot be the RSG wind because the expansion velocities are too high.
It is possible that these dense circumstellar nebulae
are produced by a very dense SN progenitor wind, as the ``deathbed
ejecta" (Chu 2001).

\begin{figure}
\plotoneb{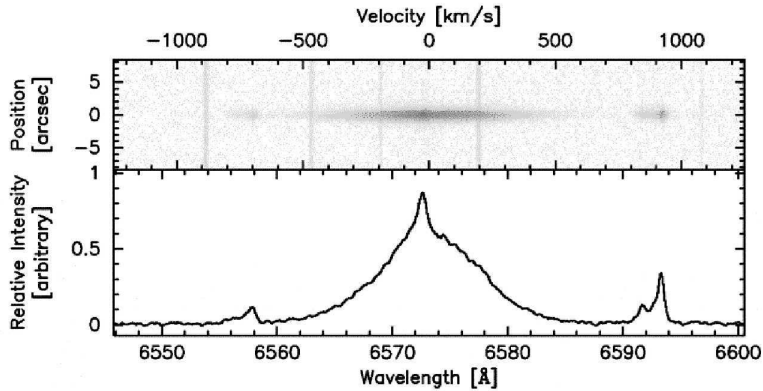}
\caption{High-dispersion spectrum of SN\,1978K at the H$\alpha$
line region.  The broad H$\alpha$ line originates from the SN
ejecta, while the narrow H$\alpha$ and [N\,{\sc ii}]
$\lambda\lambda$6548, 6583 lines originate from a circumstellar
nebula.  This figure is taken from Gruendl et al.\ (2002).}
\end{figure}

SN\,1978K (Fig.~8) shows narrow H$\alpha$ and [N\,{\sc ii}] lines 
superposed on a broad H$\alpha$ line from the SN ejecta (Chu
et al.\ 1999; Gruendl et al.\ 2002).  A lower limit on the 
nebular size can be derived from the age of the SN and the
expansion velocity of the SN ejecta, while an upper limit 
of 2.2 pc is derived from the unresolved HST WFPC2 image.
The observed auroral to nebular line ratios of the
[O\,{\sc iii}] and [N\,{\sc ii}] lines indicate a high 
density, 3-12$\times$10$^5$ cm$^{-3}$ (Ryder et al.\ 1993).
Such size and density are consistent with those of LBV nebulae. 
The progenitor of SN\,1978K might have exploded at the LBV phase.

SN\,1998S showed rapid variation in the narrow nebular line 
profiles during the first month after the SN explosion,
indicating that the SN ejecta was plowed through a circumstellar
nebula with a range of densities (Fassia et al.\ 2001).
A year after the SN explosion, a narrow [O\,{\sc iii}] line
persisted; the FWHM of this line indicates an expansion velocity
$<$25 km~s$^{-1}$ (Gruendl et al.\ 2002).  The lack of 
[N\,{\sc ii}] counterpart suggests an interstellar origin.  
Thus, this narrow [O\,{\sc iii}] line might originate from 
an interstellar bubble.

Monitoring observations of SN\,1987A have proven useful in
revealing the interaction between the SN ejecta and the
circumstellar nebula.  Likewise, monitoring observations of
extragalactic SNe with narrow nebular lines would also be 
very useful in determining the size of the circumstellar nebula.  
By comparing the density and size of a SN's circumstellar
nebula to those of ring nebular around massive stars, it
is then possible to gain insight into the nature of the
SN progenitor.


\begin{references}

\reference Aldering, G., Humphreys, R.~M., Richmond, M.\ 1994, 
  \aj\ 107, 662
\reference Bochkarev, N.~G.\ 1988, \nat\ 332, 518
\reference Brandner, W., Chu, Y.-H., Eisenhauer, F., et al.\
  1997a, \apj\ 489, L153
\reference Brandner, W., Grebel, E.~K., Chu, Y.-H., Weis, K.\ 
   1997b, \apj\ 475, L45
\reference Cappa, C.~E.\ 2002, in this volume
\reference Chu, Y.-H.\ 1981, \apj\ 249, 195
\reference Chu, Y.-H.\ 1994, in AIP Conf. Proceedings ``The Soft X-ray
  Cosmos," 154
\reference Chu, Y.-H.\ 2001, in AIP Conf. Proceedings ``Young Supernova
  Remnants," 409
\reference Chu, Y.-H., Caulet, A., Montes, M.~J., et al.\ 1999,
  \apj\ 512, L54
\reference Chu, Y.-H., Guerrero, M.~A., Gruendl, R.~A.\ 2002, in
  IAU Symposium 209 ``Planetary Nebulae," in press (astro-ph/0202509)
\reference Chu, Y.-H., Weis, K., Garnett, D.~R.\ 1999, \aj\ 117, 1433
\reference Danforth, C.~W., Chu, Y.-H.\ 2001, \apj\ 552, L155
\reference Dopita, M.~A., Bell, J.~F., Chu, Y.-H., Lozinskaya, 
   T.~A.\ 1994, \apjs\ 93, 455
\reference Fassia, A., et al.\ 2001, \mnras\ 325, 907
\reference Garc\'{\i}a-Segura, G., Langer, N., Mac Low, M.-M.\
  1996, \aap\ 316, 133
\reference Garc\'{\i}a-Segura, G., Mac Low, M.-M., Langer, N.\
  1996, \aap\ 305, 229
\reference Gervais, S., St-Louis, N.\ 1999, \aj\ 118, 2394
\reference Gruendl, R.~A., Chu, Y.-H., Dunne, B.~A., Points, S.~D.\ 
  2000, \aj\ 120, 2670
\reference Gruendl, R.~A., Chu, Y.-H., Van Dyk, S.~D., Stockdale,
   C.~J.\ 2002, \aj\ 123, 2847
\reference Humphreys, R.~M.\ 2002, in this volume
\reference Johnson, H.~M., Hogg, D.~E.\ 1965, \apj\ 142, 1033
\reference Kwok, S.\ 1983, in IAU Symposium 103 ``Planetary 
  Nebulae," 293
\reference Lamers, H.~J.~G.~L.~M., Nota, A., Panagia, N., Set al.\
  2001, \apj\ 551, 764
\reference Leitherer, C., Chavarr\'{\i}a-K., C.\ 1987, \aap\ 175, 208
\reference Marston, A.~P.\ 1997, \apj\ 475, 188
\reference Marston, A.~P.\ 2001, \apj\ 563, 875
\reference Marston, A.~P.\ 2002, in this volume
\reference Marston, A.~P., Chu, Y.-H., Garc\'{\i}a-Segura, G.\
  1994a, \apjs\ 93, 229
\reference Marston, A.~P., Welzmiller, J., Bransford, M.~A., 
  et al.\  1999, \apj\ 518, 769
\reference Marston, A.~P., Yocum, D.~R., Garc\'{\i}a-Segura, G.,
 Chu, Y.-H.\ 1994b, \apjs\ 95, 151
\reference Miller, G.~J., Chu, Y.-H.\ 1993, \apjs\ 85, 137
\reference Moore, B.~D., Hester, J.~J., Scowen, P.~A.\ 2000,
  \aj\ 119, 2991
\reference Naz\'e, Y., Chu, Y.-H., Points, S.~D., et al.\ 2001, 
  \aj\ 122, 921
\reference Nota, A., Livio, M., Clampin, M., Schulte-Ladbeck, R.\
   1995, \apj\ 448, 788
\reference Nota, A.\ et al.\ 1996, \apjs\ 102, 383
\reference Nota, A., Smith, L., Pasquali, A., Clampin, M., Stroud, M.\
  1997. \apj\ 486, 338
\reference Pasquali, A., Nota, A., Clampin, M.\ 1999, \aap\ 343, 536
\reference Pasquali, A., Nota, A., Langer, N., et al.\
   2000, \aj\ 119, 1352
\reference Pasquali, A., Schmutz, W., Nota, A., Origlia, L.\ 1997,
 \aap\ 327, 265
\reference Pittard, J.~M., Dyson, J.~E., Hartquist, T.~W.\
   2001a, \aap\ 367, 1000
\reference Pittard, J.~M., Hartquist, T.~W., Dyson, J.~E.\ 
   2001b, \aap\ 373, 1043
\reference Podsiadlowski, P., Joss, P.~C., Hsu, J.~J.~L.\ 1992,
   \apj\ 391, 246 
\reference Rizzo, J.~R., Mart\'{\i}n-Pintado, J., Desmurs, J.-F.\ 2002,
  in this volume
\reference Rizzo, J.~R., Mart\'{\i}n-Pintado, J., Henkel, C.\ 2001
  \apj\ 553, 181
\reference Ryder, S., Staveley-Smith, L., Dopita, M.~A., et al.\
  1993, \apj\ 416, 167
\reference Salamanca, I., et al.\ 1998, \mnras\ 300, L17
\reference Salamanca, I., Terlevich, R., Tenorio-Tagle, G.\ 2002,
  \mnras\ 330, 844
\reference Smartt, S.~J.\ 2002, in this volume
\reference Smartt, S.~J., Gilmore, G.~F., Tout, C.~A., Hodgkin,
 S.~T.\ 2002, \apj\ 565, 1089
\reference Smartt, S.~J., Gilmore, G.~F., Trentham, N., et al.\
  2001, \apj\ 556, L29
\reference Smith, L.~J., Nota, A., Pasquali, A., Leitherer, C., et
  al.\ 1998, \apj\ 503, 278
\reference Smith, N., Humphreys, R.~M., Davidson, K.\ et al.\ 
  2001, \aj\ 121, 1111
\reference Smith, N., Gehrz, R.~D., Humphreys, R.~M., et al.\
  1999, \aj\ 118, 960
\reference St-Louis, N., Doyon, R., Chagnon, F., Nadeau, D.\ 1998,
  \aj\ 115, 2475
\reference Walborn, N.~R.\ 1972, \aj\ 77, 312
\reference Walborn, N.~R., et al.\ 1989, \aap\ 219, 229
\reference Weaver, R., McCray, R., Castor, J., Shapiro, P., Moore, R.\ 
  1977, \apj\ 218, 377
\reference Weis, K.\ 2002, in this volume
\reference Wrigge, M.\ 1999, \aap\ 343, 599
\reference Wrigge, M., Chu, Y.-H., Magnier, E.~A., Kamata, Y.\
  1998, in Lecture Notes in Physics, vol.506, 425
\reference Wrigge, M., Wendker, H.~J., Wisotzki, L.\ 1994,
  \aap\ 286, 219

\end{references}
\end{document}